\documentclass[acus]{JAC2000}

\usepackage{graphicx}

\begin{document}

\title{\flushright{WECT006}\\[15pt] \centering 
BEAM FEEDBACK SYSTEMS AND BPM READ-OUT SYSTEM\\
FOR THE TWO-BUNCH ACCELERATION AT THE KEKB LINAC}

\author{K. Furukawa\thanks{e-Mail: \texttt{<kazuro.furukawa@kek.jp>}},
N. Kamikubota, T. Suwada\\ 
High Energy Accelerator Research Organization (KEK),
Tsukuba, Ibaraki, 305-0801, Japan\\
T. Obata\\
Mitsubishi Electric System Service, Tsukuba, Ibaraki, 305-0045, Japan
}

\maketitle

\begin{abstract}

In order to double the positron injection rate into the KEKB ring, a
two-bunch acceleration scheme has been studied at the linac, in which
bunches separated by 96 ns are accelerated in 50 Hz.  In this scheme
stabilization of the energy and orbit of each bunch is indispensable.
Thus, the beam energy and orbit feedback systems have been upgraded.

Since beam characteristics are acquired through beam-position monitors
(BPM), their read-out system was improved to meet two-bunch
requirements.  Combined waveforms from BPM's were adjusted with delay
cables avoiding overlaps so as to enable the simultaneous measurements of
the beam positions of two bunches.

The beam energies of two bunches were balanced by tuning the rf pulse
timings, and the average energy was stabilized by adjusting the
accelerating rf phases. The average beam orbits were also stabilized.
Slow feedback systems at the injector section for charge and bunching
stabilities are being planned as well.  These systems were 
successfully used in the test beams and will be employed during routine
operation.

\end{abstract}

\section{INTRODUCTION}

The electron/positron linac at KEK injects 8-GeV electron and 3.5-GeV
positron beams into KEKB rings, where the CP-violation study is
carried out. Since the efficiency of the experiment can be increased by
shortening the injection time, several mechanisms have been introduced
to accomplish this\cite{bi-switch-fb-lin2000,co-efb-ical99}.  Especially,
much effort has been made to improve the positron injection time,
since it is longer compared with that of electrons\cite{lin-com-linac00}.

One of such effort is a two-bunch acceleration plan, which has been 
studied and applied\cite{ohsawa-twobunch-pac01,ogawa-twobunch-apac01}. 
In this scheme two bunches of
positrons are accelerated in one rf pulse, which is 50 Hz; they may
double the injection rate.  The time space between two
bunches, however, is restricted by the rf frequencies of the linac 
and the rings, 
and the smallest space is 96.29 ns, since the common frequency is 
10.38 MHz.  Thus, a precise beam control and 
diagnosis are necessary. 

The beam diagnosis used so far has been made by employing strip-line-type
beam-position monitors (BPM), wire scanners for transverse profiles and 
streak cameras for longitudinal profiles.  In order to maintain 
stable beams, it is essential to have the beam instrumentations work 
for both of the two bunches.  The two-bunch read-out of BPMs is especially
important, because it is used in a number of orbit and energy feedback
loops to stabilize the beams.

\section{BPM AND READ-OUT SYSTEM}

\begin{figure}[bt]
 \centering
 \includegraphics*[width=70mm]{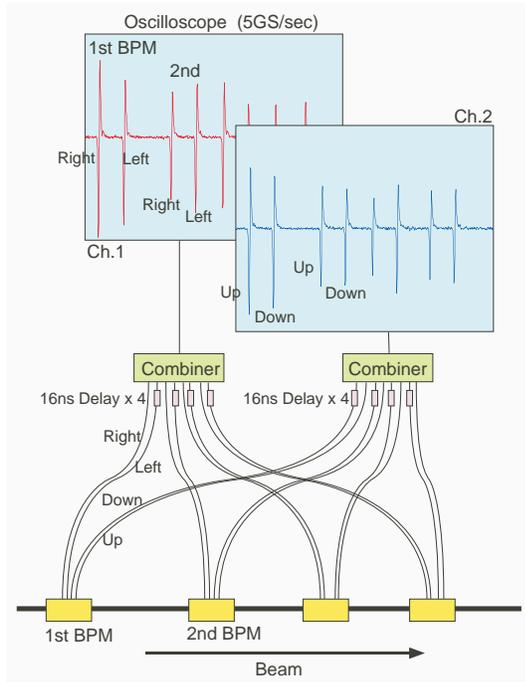}
 \caption{Signals from multiple BPMs arranged so that 
  combined waveforms can be processed properly.}
 \label{fig1}
\end{figure}

Along the 600-m linac, 90 BPMs are installed and their signals are 
transferred to one of 18 measurement stations.  Signals are 
delayed and combined so as not to overlap each other, and are fed into 
a 5-Gs/s waveform digitizer 
(Sony-Tektronix TDS-680B/C)\cite{bpm-system-nim}, as in Fig.~\ref{fig1}. 
Although the BPM signal is a fast bipolar, the readout precision 
is optimized using the interpolation function of the digitizer. 
All 18 digitizers are triggered by a single distributed signal, which is 
synchronized with beam repetition and rf frequencies. 

The waveform is read through the GPIB, and a signal from each electrode 
is analyzed with a predetermined response function once per second by 
a VME computer (Force 68060). 
The response functions include 3rd-order position-mapping functions, 
attenuation factors of various components and position offsets against 
the center of the corresponding quadrupole magnet derived from a beam-based 
alignment. 

Since the timing and amplitude ranges of BPM signals are different 
depending on the beam modes and locations, the process is driven by 
a control database system\cite{bpm-daq-ical99}. 

The acquired beam positions at 18 stations are sent to central computers 
once per second 
and are served for various beam-energy and orbit feedback systems to 
maintain stable beam operation. 

\section{ TWO-BUNCH OPERATION }

The BPM system was improved for two-bunch operation. 

\subsection{ Improvements to BPM System }

As written above, it is important to acquire the beam positions of 
two bunches along the linac simultaneously to study the beams.  
In our instrumentation, signals from those two bunches appear 
as two signals separated by 96.29 ns on the waveform.  Although 
it was sometimes necessary to add more delay lines so as to avoid 
waveform overlapping, there was no need to add any specific hardware 
to handle such signals with small separations.  

The calibration factors were re-examined since delay lines were added, 
and the beam-timing database for the signal analysis was extended to 
accommodate two-bunch information.  

Processing functions/commands for BPMs on the central 
computers are also extended or added for two bunches, while keeping 
the old functions as before for single-bunch operations.  

With these modifications, the BPM processing system was extended 
for two-bunch operation without any performance loss in either
precision and speed.  It has been used in beam operation 
since March, 2001. 

\subsection{ Operation Software with BPM }

\begin{figure}[bt]
 \centering
 \includegraphics*[width=65mm]{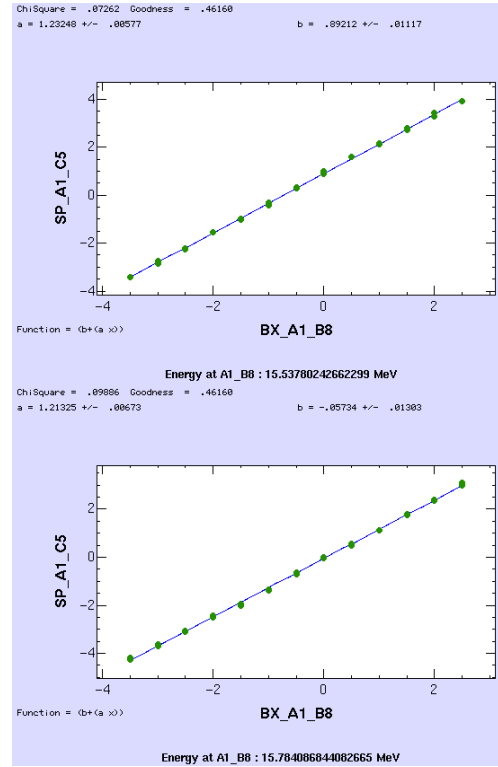}
 \caption{Software panel to measure the beam energies of 
  two bunches with a steering magnet (BX\_A1\_B8) and a BPM 
  (SP\_A1\_C5).  After adjusting the timing of the electron gun, they became 
  almost the same energy 15.5 MeV and 15.8 MeV.}
 \label{fig2}
\end{figure}

Most of the operation software which utilize the BPM information was 
extended to meet both single- and two-bunch operations.  One of 
such examples is Fig.~\ref{fig2}, which measures the beam energies of 
two bunches by correlation between a steering-magnet field and 
the beam-position response at the bunching section.  

\subsection{ Two Bunch Controls }

In order to accelerate the beams properly, the beam characteristics 
of two bunches need to be adjusted so as to be the same.  For example, 
in order to adjust the beam-energy differences, we change the 
beam timing and rf pulse timing.  The beam timing can be changed 
by 10-ps steps\cite{ohsawa-twobunch-pac01} and the rf pulse 
timing can be changed by 1.75-ns steps at each sector independently. 
Most of other parameters in the linac are not sensitive against  
time separation of 96.29 ns. 



\begin{figure}[bt]
 \centering
 \includegraphics*[width=80mm]{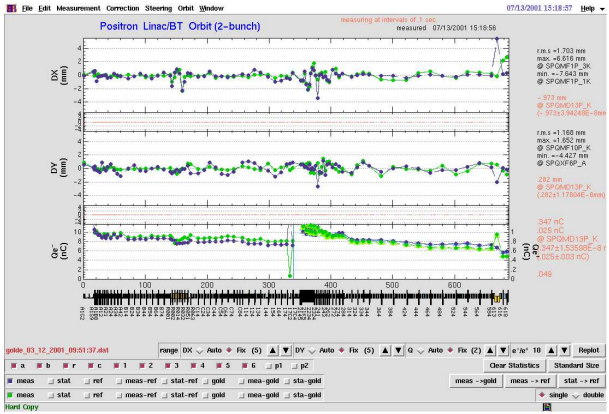}
 \caption{Beam orbit and charge of electrons (left) and positrons 
  (right) along the 600-m linac.  Two slightly different lines 
  indicate the first and second bunches.}
 \label{fig3}
\end{figure}

With such adjustments, the 10-nC primary electron bunches are 
accelerated up to 3.7 GeV and positrons are generated as shown in 
Fig.~\ref{fig3}. 

\subsection{ BEAM FEEDBACK LOOPS }

The beam feedback loops in the linac for energy and orbit 
stabilization\cite{co-efb-ical99}
were also extended to control two-bunch beams.  Since we don't have 
many mechanisms to control two bunches independently, most feedback 
loops were modified to use positions derived from the charge-weighted 
averages of two bunches.  With these changes, those loops can maintain 
the average orbit and energy.  In software, only the monitoring function 
was extended to read the average positions if two bunches are accelerated. 
For positron injection, about 20 beam feedback loops are used, and 
they are all extended for two bunches. 

\begin{figure}[bt]
 \centering
 \includegraphics*[width=80mm]{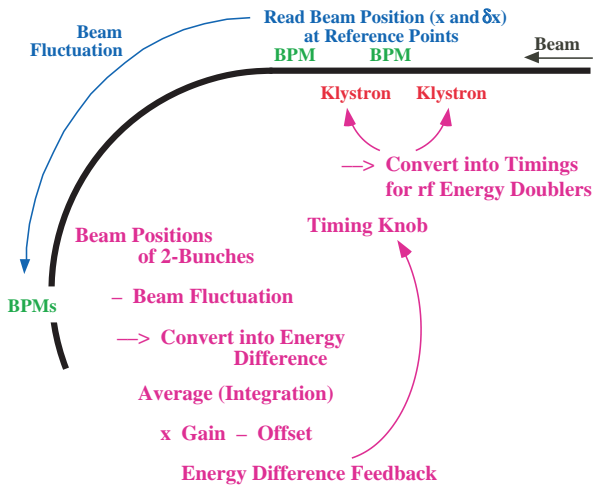}
 \caption{Energy-difference feedback loop at the J-arc.}
 \label{fig4}
\end{figure}

While normal energy and orbit feedback loops use charge-weighted 
average positions, feedback loops to minimize the energy differences 
use the position difference between two bunches, as shown in 
Fig.~\ref{fig4}. 

Although the energy difference does not change frequently, such 
loops stabilize the beam over the long term. 

\section{SUMMARY}

The data-acquisition system for the linac BPMs was upgraded to 
provide beam positions in two-bunch operation without losing any 
original features.  Along with improvements of the streak camera and 
wire scanner systems, it has still been indispensable to study and 
operate on linac beams.  The system is also used by many operation 
software programs, including beam-energy and orbit feedback systems.


\begin{thebibliography}{9}

\bibitem{bi-switch-fb-lin2000}
K.~Furukawa {\it et~al.},
``Beam Switching and Beam Feedback Systems at KEKB Linac'',
{\em Proc. of LINAC2000}, Monterey, USA., 2000, p.633.

\bibitem{co-efb-ical99}
K.~Furukawa {\it et~al.},
``Energy Feedback Systems at KEKB Injector Linac'',
{\em Proc. of ICALEPCS99}, Trieste, Italy, 1999, p.248.

\bibitem{lin-com-linac00}
K.~Furukawa {\it et~al.},
``Towards Reliable Acceleration of High-Energy and High-Intensity 
Electron Beams'', 
{\em Proc. of LINAC2000}, Monterey, USA., 2000, p.630.

\bibitem{ohsawa-twobunch-pac01}
S.~Ohsawa {\it et~al.},
``Increase of Positrons by High-intensity Two-bunch Acceleration
Scheme at the KEKB Linac'', to be published in 
{\em Proc. of PAC2001}, Chicago, USA., 2001.

\bibitem{ogawa-twobunch-apac01}
Y.~Ogawa {\it et~al.},
``Two-Bunch Operation of the KEKB Linac for Doubling
  the Positron Injection Rate to the KEKB Ring'', to be published in 
{\em Proc. of APAC2001}, Beijing, China, 2001.

\bibitem{bpm-system-nim}
T.~Suwada {\it et~al.},
``Stripline-Type Beam-Position-Monitor System
for Single-Bunch Electron/Positron Beams'',
Nucl. Instr. and Meth. {\bf A440} (2000) 307.

\bibitem{bpm-daq-ical99}
N.~Kamikubota {\it et~al.}, 
``Data Acquisition of Beam-Position Monitors for the KEKB Injector-Linac'',
{\em Proc. of ICALEPCS99}, Trieste, Italy, 1999, p.217.

\end{thebibliography}
\end{document}